\documentclass[fleqn,twoside]{article}
\usepackage{espcrc2}
\usepackage{graphicx}
\input epsf

\newcommand{\frat}[2]{\frac{\textstyle #1}{\textstyle #2}}

\newcommand{\vf}[1]{\mbox{\boldmath $#1$}}

\title{Instanton vacuum at finite density of quark matter}

\author{S.V. Molodtsov\address[M]{State Research Center,
Institute of Theoretical and Experimental Physics,\\
117259, Moscow, Russia},
        G.M. Zinovjev\address[Z]
{Bogolyubov Institute for Theoretical Physics,\\
UA-03143, Kiev, Ukraine}}

\begin{document}

\begin{abstract}
  We study light quark interactions in the instanton liquid at finite
quark/baryon number density analyzing chiral  and  diquark
condensates and  investigate the behaviors of quark dynamical mass
and both condensates together with instanton liquid density as a
function of quark chemical potential. We conclude the quark impact
(estimated in the tadpole approximation) on the instanton liquid
could shift color superconducting phase transition to higher
values of the chemical potential bringing critical quark matter
density to the values essentially higher than conventional nuclear
one. \vspace{1pc}
\end{abstract}

\maketitle

  Aiming to provide a good insight into the real world we are
interested in deriving the QCD phase diagram from the first
principles as a function of temperature, chemical potential and
number of flavors. While much was has been learned about the phase
structure of QCD at finite temperature (combining the perturbative
theory analysis and lattice simulations), this structure at
nonzero quark/baryonic densities has been less explored. The
former discovered that with the temperature dropped below the
critical value, chiral symmetry is broken and quarks become
confined. Moreover, both fundamental phenomena, seems, take place
at the same temperature indicating they are tightly knitted
\cite{karsh}. This result is considered as one of the most
indicative to investigate the QCD dynamics. Indeed, the lattice
studies allow us to compare the changing behavior of physical
observables on both $T_c$ sides providing very important
information on the underlying excitations responsible for
confinement or chiral symmetry breaking. And if the letters are
more understandable being, according the common wisdom, the
instanton-like excitations, different ideas for those responsible
for confinement are still under debates. We believe the recent
developments \cite{we} could be illuminating on the way of
resolving this fundamental problem.
\begin{figure}[htb]
\includegraphics[scale=0.38]{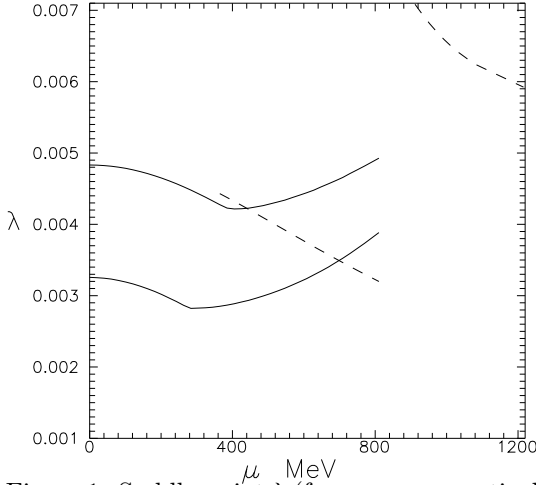}
\vspace{-1.2cm}
 \caption{Suddle-point $\lambda$ (free energy,
practically) as the function of chemical potential $\mu$. The
upper solid and lower dashed lines correspond to the solutions
with zero diquark condensate and zero chiral condensate,
respectively. The lower solid and upper dashed lines correspond to
the same solutions but including the quark interaction with IL
when the modified functional integral is evaluated in tadpole
approximation.} \label{fig1}
\end{figure}

The factorized form of the generating functional of instanton
liquid (IL) model \cite{1}, \cite{2} ${\cal Z}={\cal
Z}_g\cdot{\cal Z}_\psi$, where the factor ${\cal Z}_{g}$ provides
nontrivial gluon condensate while the fermion part ${\cal
Z}_{\psi}$ is responsible to describe the chiral and diquark
condensates in instanton medium, allows us to investigate the
condensate excitations. ${\cal Z}_{g}$ is supposed to be saturated
by the superposition of the pseudo-particle (PP) fields
((anti-)instantons).

The quark fields are considered to be {\it influenced} by the
certain stochastic ensemble of PPs while calculating the quark
determinant. The fermion field action is approached by the zero
modes which are the solutions of the Dirac equation  $[i\hat
D(A_{\bar I I})-i\mu \gamma_4]~\Phi_{\bar I I}=0$ with chemical
potential $\mu$ \cite{abr}, in the field of PP. With the auxiliary
integration over the parameter $\lambda$ the quark determinant
$Z_\psi$ may be exponentiated to the following form with
dimensionless variables for the massless quarks of two flavors and
three colors
\begin{eqnarray}\label{d1}
&&\!\!\!\!\!\!\!\!\!\!\!{\cal Z}_\psi\sim\int\!\! d\lambda\int\!\!
D\psi^\dagger D\psi\exp\left\{
N~\left(\ln\frat{n\bar\rho^4}{\lambda R}-1\right)\right\}
\!\!\times\nonumber\\
&&\!\!\!\!\!\!\!\!\!\times\exp\left\{\int
\frat{dk}{\pi^4}\sum_{f=1}^{2}\psi^\dagger_{f}(k) (-\hat
k-i\hat\mu)\psi_{f}(k)+V\right\},\nonumber
\\
&&\!\!\!\!\!\!\!\!\!V=V_L+V_R,~V_{L}=\lambda~(\psi^{\dagger L}_1
L_1\psi_1^{L}) (\psi^{\dagger L}_2 L_2\psi_2^{L}),\nonumber
\end{eqnarray}
and to get the right hand chiral components one should change
$L\to R$, $\psi_f^{T}=(\psi_f^{R},\psi_f^{L})$, and interaction
terms are defined by the corresponding Fourier-components of zero
modes \cite{diakcar}, \cite{rapp}, $\mu_\nu=({\vf 0},\mu)$. Then
we are well armed to calculate the chiral condensate
$$\langle\psi^\dagger (k)\psi(l)\rangle=
-\pi^4 \delta(k-l)~Tr~S(k)~,$$ if the quark Green function $S(k)$
is known and the diquark condensate also
$$
\langle\psi^{L,R}_{1\alpha i}(k)\psi^{L,R}_{2\beta j}(l)\rangle=
\epsilon_{12}~\epsilon_{\alpha\beta}~\pi^4~\delta(k+l)~F^{L,R}_{ij}(k)~,
$$
as the solutions of the  Gorkov-Dyson-Schwinger equations.
\begin{figure}[htb]
\includegraphics[scale=0.38]{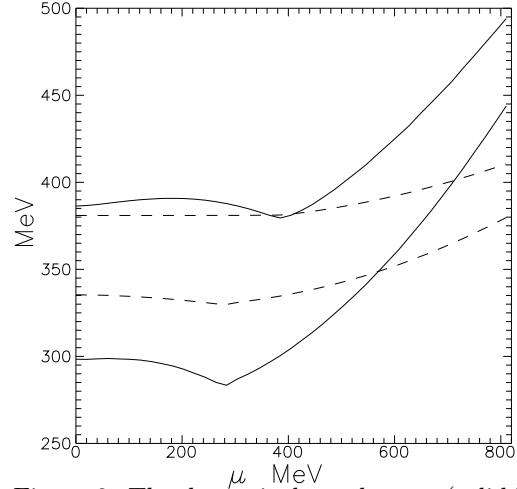}
\vspace{-1.2cm} \caption{The dynamical quark mass (solid lines)
and chiral condensates (dashed lines) as function of chemical
potential $\mu$. The lower solid and dashed curves correspond to
the calculation with the modified generating functional.}
\label{fig2}
\end{figure}
 The corresponding solutions for the suddle point
$\lambda$ of ${\cal Z_\psi}$ are shown in Fig. \ref{fig1}
(practically free energy in one loop approximation). The quark
feedback upon the instanton background is pretty limp and could be
perturbatively incorporated as a small variation of instanton
liquid parameters $\delta n$ and $\delta \rho$ around their
equilibrium values of $n$ and $\bar\rho$. We describe further the
{\it quark feedback} dealing with PPs changing their sizes
adiabatically i.e. $\rho\to\rho(x,z)$. It results to the
interaction of quarks and scalar field of the $\delta\rho$
deformation what in turn modifies the $L$ and $R$ kernels in  the
generating functional. The corresponding solutions of the modified
(in tadpole approximation) suddle point equations are also exposed
in Fig. \ref{fig1}. The dynamical quark mass $M$ (solid lines) and
the quark condensate
$-i\langle\psi^\dagger\psi\rangle=i~Tr~S(x)|_{x=0}$ (dashed lines)
are presented in Fig. \ref{fig2} where the lower curves correspond
to the calculation with modified generating functional. Through
this paper the value of renormalization constant is fixed by
$\Lambda=280~{\mbox{MeV}}$. Fig. 3 shows the instanton liquid
density for the phase with non-zero chiral condensate (diquark
condensate absent). Let us notice that all the curves give the
reliable estimates in the interval extending till the vicinity of
crossing points.

\begin{figure}[htb]
\includegraphics[scale=0.38]{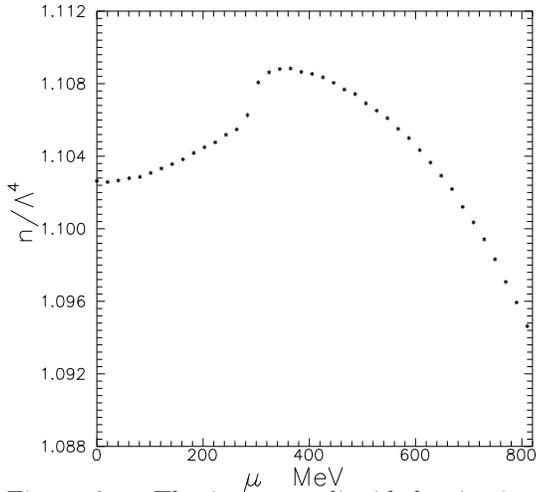}
\vspace{-1.2cm} \caption{ The instanton liquid density in the
phase of broken chiral symmetry.} \label{fig3}
\end{figure}

\begin{figure}[htb]
\includegraphics[scale=0.38]{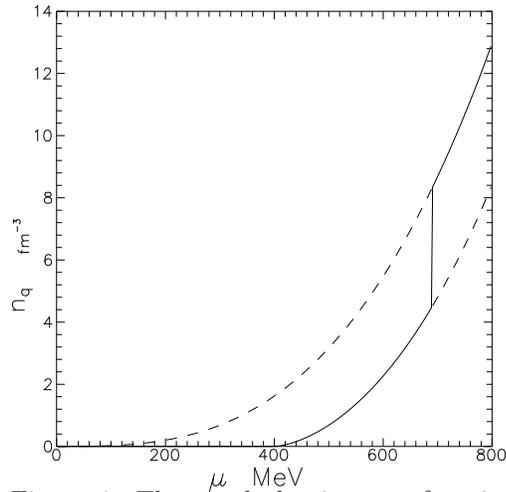}
\vspace{-1.2cm} \caption{The quark density as a function of $\mu$
reconstructed  from the branch of solid curve till the second
$\mu_c$ and branch of dashed one beyond $\mu_c$. The solid line
corresponds to the stable phase, dashed curve corresponds to the
metastable one.} \label{fig4}
\end{figure}

Apparently, the physical picture which might be extracted from the
results obtained  is regulated by the crossing points of solid and
dashed curves in Fig. \ref{fig1}. That corresponding the smallest
critical value of the chemical potential $\mu_c\simeq 400~
{\mbox{MeV}}$ (for the nonperturbated IL), in principle, is in
conformity with the estimate given in \cite{diakcar}, \cite{rapp}
($\mu_c\simeq {\mbox{340}}~MeV$). Meanwhile, in \cite{diakcar} it
was mentioned that such an estimate leads to nonrealistic values
of critical quark density for the transition to  color
superconducting phase. That estimate occurs to be quite comparable
with conventional nuclear density $n_0=0.45~{\mbox {fm}}^{-3}$.
Our approach teaches including the quark interaction with IL
certainly increases  $\mu_c$. It is clear already from the second
$\mu_c\simeq 700~{\mbox{MeV}}$ (the corresponding quark density is
exhibited in Fig. \ref{fig4}). Moreover, the position of upper
dashed curve in Fig. \ref{fig1} (the perturbated IL) signals that
the $\mu_c$ for transition to color superconducting phase might be
shifted to very high values laying, rather, beyond the validity
interval of IL model. It happens because the interquark Coulomb
field strengths are comparable or even exceed the instanton ones
and therefore the assumption about the saturating configurations
becomes invalid.

\end{document}